# Q-NAV: 수중 무선 네트워크에서 강화학습 기반의 NAV 설정 방법

박석현[1], 조오현[2*]
[1]충북대학교 소프트웨어학과 학생, [2]충북대학교 소프트웨어학과 교수

# Q-NAV: NAV Setting Method based on Reinforcement Learning in Underwater Wireless Networks

Seok-Hyeon Park[1], Ohyun Jo[2*]
[1]Student, Department of Computer Science, ChungBuk National University
[2]Professor, Department of Computer Science, ChungBuk National University

**요 약** 수중 자원 탐색 및 해양 탐사, 환경 조사 등 수중 통신에 대한 수요가 급격하게 증가하고 있다. 하지만 수중 무선 통신을 사용하기 앞서 많은 문제점을 가지고 있다. 특히 수중 무선 네트워크에서 환경적 요인으로 인해 불가피하게 발생하는 불필요한 지연 시간과 노드 거리에 따른 공간적 불평등 문제가 존재한다. 본 논문은 이러한 문제를 해결하기 위해 ALOHA-Q를 기반으로 한 새로운 NAV 설정 방법을 제안한다. 제안 방법은 NAV 값을 랜덤하게 사용하고 통신 성공, 실패 유무에 따라 보상을 측정한다. 이후 보상 값에 따라 NAV 값을 설정한다. 수중 무선 네트워크에서 에너지와 컴퓨팅 자원을 최대한 낮게 사용하면서 NAV 값을 강화 학습을 통하여 학습하고 한다. 시뮬레이션 결과 NAV 값이 해당 환경에 적응하고 최선의 값을 선택하여 불필요한 지연 시간문제와 공간적 불평등 문제를 해결할 수 있음을 보여준다. 시뮬레이션 결과 설정한 환경 내에서 기존 NAV 설정 시간 대비 약 17.5%의 시간을 감소하는 것을 보여준다.

**주제어 :** 강화학습, Q-learning, ALOHA-Q, 수중 무선 네트워크, 수중 IoT 네트워크, 센서, RTS, CTS, NAV

**Abstract** The demand on the underwater communications is extremely increasing in searching for underwater resources, marine expedition, or environmental researches, yet there are many problems with the wireless communications because of the characteristics of the underwater environments. Especially, with the underwater wireless networks, there happen inevitable delay time and spacial inequality due to the distances between the nodes. To solve these problems, this paper suggests a new solution based on ALOHA-Q. The suggested method use random NAV value. and Environments take reward through communications success or fail. After then, The environments setting NAV value from reward. This model minimizes usage of energy and computing resources under the underwater wireless networks, and learns and setting NAV values through intense learning. The results of the simulations show that NAV values can be environmentally adopted and select best value to the circumstances, so the problems which are unnecessary delay times and spacial inequality can be solved. Result of simulations, NAV time decreasing 17.5% compared with original NAV.

This research was a part of the project titled 'Development of Distributed Underwater Monitoring and Control Networks', funded by the Ministry of Oceans and Fisheries, Korea. And this work was supported by the research grant of the Chungbuk National University in 2018.
*Corresponding Author : Ohyun Jo(Ohyunjo@chungbuk.ac.kr)
Received November 26, 2020    Revised November 26, 2020
Accepted January 20, 2020     Published January 28, 2020

# 1. 서론

## 1.1 서론

지구 표면의 71%는 물로 구성되어 있다[1]. 대부분의 국가는 해안 지역을 포함하고 있으며, 해안 자원 및 연관 산업을 통해 다양한 수익 창출 및 기술 개발이 이루어지고 있다. 최근에는 지상에서 수집되는 에너지 및 자원 고갈의 문제를 해결하기 위해 수중 자원에 대한 탐색 및 연구가 활발하게 진행되고 있다. 또한 수중 오염 문제 해결과 지진, 해일, 태풍과 같은 자연재해 분석 및 예측을 위한 기술 개발, 해안 경비 및 해양 국방 산업, 어업 등 다양한 기술 개발 수요가 존재한다[2,3].

이러한 기술은 주로 수중에서 데이터를 습득하여 활용하게 된다. 이때, 수중 데이터를 수집하고 활용하기 위해서는 수집을 위한 통신 시스템이 필요하다. 수중 통신의 경우 유선 또는 무선 통신 시스템을 활용할 수 있다. 최근의 통신 장비들은 주로 무선 통신 시스템을 도입하고 있다. 수중 통신 특성상 통신 범위가 광범위한 문제와 배, 잠수함, 수중 생물, 잠수부 등 수중 이동 매체가 존재하는 문제 때문에 유선 통신을 설치하는 것은 현실적으로 힘들다. 또한 AUV (Autonomous underwater vehicle)나 ROV (Remotely Operated Underwater Vehicle), 잠수함 등 데이터를 수집하는 노드가 이동성을 갖는 경우가 존재하는데, 이러한 이동성을 갖는 노드에 대해서는 더욱 유선 통신 시스템을 활용하기 어렵다. 위와 같은 이유로 수중에서는 유선 네트워크 환경을 사용이 힘들어 무선 네트워크 환경 적용되고 있으며, 이에 대한 다양한 연구가 이루어지고 있다.

하지만 수중 무선 네트워크 또한 환경적인 한계점을 가지고 있다. 수중 무선 네트워크의 경우 환경 특성상 RF 신호의 활용이 어려워 주로 음파를 사용한다. 수중에서 음파를 사용하는 경우 느린 전파 속도(약 1500m/s)와 선박 엔진, 스크루 소음 등 잡음에 대해 취약하다는 문제, 전송 효율 대비 높은 에너지 소모, 주로 배터리를 사용함에 있어 에너지 제한적인 조건 등 고려해야 할 많은 문제점을 가지고 있다[3–5]. 특히 해결해야 하는 주요 문제점 중 하나는 느린 전파 속도와 느린 데이터 처리 속도, 최적화되지 못한 통신대기 시간이 일으키는 전체 네트워크 성능 저하이다. 수중 무선 네트워크의 경우 충돌을 방지하기 위해 해당 환경에서의 최대 전파 지연 시간을 이용하는데, 이때 발생하는 불필요한 대기시간이 존재한다. 이는 GPS와 같은 장비가 사용 불가하여 노드 간 위치를 알 수 없기 때문에 발생한다. 수중 무선 통신 환경에서 RTS/CTS를 사용하기 위해 NAV 값을 계산한다. 하지만 수중 네트워크에서는 노드 간 위치를 모르는 문제로 인하여 최대 전파 지연시간을 이용하게 되고 이는 불필요한 지연 시간이 생기게 되는 것이다.

이를 최적화하기 위해서는 통신을 이용하여 노드 간 위치를 계산하여야 한다. 하지만 계산량이 높고 에너지 소모가 많기 때문에 배터리를 사용하는 수중 통신에서는 이와 같은 방법을 적용하기 어렵다. 이러한 문제를 해결하고자 하는 연구가 일부 진행되었으나 프레임 내 새로운 필드를 추가해야하는 단점을 갖고 있다[6, 7]. 정해진 통신 프로토콜을 수정해야하는 점은 모든 통신에 적용하기 힘들 수 있다는 치명적이다.

본 논문은 수중 무선 네트워크 성능을 향상시키기 위해 불필요한 NAV(Network Allocation Vector) 대기시간을 제거하고, 최선의 값을 선택 한다. NAV 대기시간 선택 방법은 Q-learning에서 미래 보상에 대한 개념을 제외한 ALOHA-Q를 사용한다[8–10]. ALOHA-Q는 환경에 따라 간단한 연산으로 최적의 학습이 가능하다. 이는 에너지 제한적이고 낮은 컴퓨팅 성능을 갖는 수중 무선 네트워크 환경에서 적은 에너지, 복잡도가 낮은 알고리즘을 이용, 좋은 성능을 이끌어내기 최적화되어있으며, 네트워크의 전반적인 성능 향상을 일으킬 것으로 예상된다.

### 1.1.1 RTS/CTS/NAV

RTS/CTS (Request to Send / Clear to Send) 는 IEEE 802.11 무선 네트워크 프로토콜에서 상황에 따라 선택적으로 사용할 수 있는 통신 기술이다[11]. RTS/CTS는 채널 예약, 즉 프레임 충돌 문제를 막기 위해 사용되며, 은닉 노드 문제(hidden node problem)를 해결하기 위해 개발되었다. RTS/CTS는 송신 측이 데이터 전송을 위해 무선 링크를 예약한다. 이 때, 사용되는 프레임을 RTS/CTS라 한다. RTS/CTS의 작동 방식은 Fig. 1.과 같다.

Fig. 1.에서는 Node 1이 Sink Node에게 데이터를 전송하기 위해 채널을 예약하는 과정을 보여준다. Node 1은 RTS 프레임을 주변 노드에 전송하여 채널을 예약한다. RTS를 수신한 다른 노드들은 NAV 값을 설정하고 설정된 시간만큼 채널이 점유되었다고 판단하여 송출 대기 상태에 들어간다.

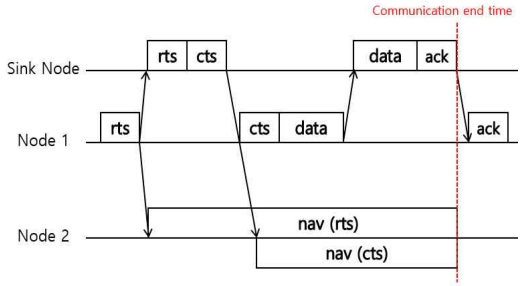

Fig. 1. RTS/CTS operations

RTS를 수신한 Sink Node의 경우 데이터 수신 준비가 되었음을 알리는 CTS 프레임을 전송한다. CTS를 수신한 다른 노드는 RTS 수신 때와 마찬가지로 NAV 값 설정 후 송출 대기 상태에 들어간다. NAV 값은 타이머와 같은 개념이다. NAV 값이 0되는 순간부터 데이터 송출을 위한 경쟁이 가능하다. NAV 값의 경우 송신측에서 채널 점유 소요 예상 시간을 계산하고 RTS/CTS 프레임 내 정해진 필드에 넣어 전송한다. 수신측은 해당 프레임 내부 필드를 확인하여 NAV를 설정하게 된다.

## 2. 문제점 및 제안 방법
### 2.1. 수중 무선 네트워크에서의 문제점

본 연구에서는 수중 무선 네트워크에서 충돌 방지를 위해 RTS/CTS를 사용함에 있어 발생하는 문제점을 해결하고자 한다. 무선 LAN 802.11에서 옵션으로 사용되는 RTS/CTS의 경우 수중 무선 네트워크에 적용하여 사용하는 경우 크게 두 가지 문제가 발생한다. 첫 번째로는 불필요한 대기 시간 발생 문제, 두 번째는 노드 간 경쟁 불공정성 문제가 발생한다. 본 장에서는 두 가지 문제점에 대한 설명과 이를 해결하기 위한 방법을 제안한다.

#### 2.1.1. NAV 대기시간

수중 무선 네트워크에서 RTS/CTS 옵션을 사용하는 경우 불필요한 대기시간이 발생되는 문제가 존재한다. 이러한 문제는 네트워크의 전체적인 처리량을 저하시킨다. 수중 무선 네트워크에서 NAV 값을 설정하는 방법은 각각 다음과 같다.

$$NAV_{RTS} = 3*Propagation\ delay\ time + CTS + DATA + ACK \quad (1)$$

$$NAV_{CTS} = 2*Propagation\ delay\ time + DATA + ACK \quad (2)$$

수중 무선 통신의 경우 GPS를 사용할 수 없어 노드 간 위치를 알 수 없다. 이는 NAV 값을 계산할 때, 노드 사이의 지연시간을 알 수 없음을 의미한다. 이러한 문제로 인하여 수중 무선 네트워크에서는 NAV 설정 시 사용되는 Propagation delay time을 해당 환경에서의 최대 전파 지연시간으로 설정하여 이용할 수밖에 없다. 이는 가까운 노드 간 통신에서도 최대 전파 지연시간을 이용하여 NAV 값을 설정하기 때문에 불필요한 대기 시간을 초래한다.

#### 2.1.2. 노드 간 불공정성 문제

수중 무선 네트워크에서 RTS/CTS 옵션을 사용하는 경우, 노드 간 통신 불공정성 문제가 발생한다. 이러한 문제는 Propagation delay time을 최대 전파 지연시간으로 사용함에 있어 발생한다. 수중 무선 네트워크 환경에서는 노드 간 거리에 상관없이 최대 전파 지연시간을 사용하여 동일한 NAV 값을 가지게 된다.

송신 노드로부터 거리가 먼 노드는 거리가 가까운 노드에 비해 더 늦게 RTS/CTS 패킷을 수신하게 된다. 특히 음파를 사용하는 수중 무선 통신의 경우 전송 속도가 느리기 때문에 거리에 따른 수신 시각 차이가 크다. 각 노드가 RTS/CTS 수신 시각은 다르나 NAV 설정 시간이 같아 송신 측으로부터 거리가 먼 노드는 가까운 노드 대비 대기 시간이 더 늦게 끝나게 된다.

이러한 시간 차이로 인해 데이터 송신 기회에 있어 거리가 먼 노드는 상대적으로 불리하게 된다. 또한 NAV 대기 시간 이후에 데이터 전송이 가능해진 상황에서 채널 예약을 하려고 하는 경우, 예약을 위한 RTS 패킷이 수신 측으로부터 가까운 노드 대비 수신 노드까지의 전송 시간이 더 오래 걸리기 때문에 거리에 따른 불공정성 문제는 커질 수밖에 없다.

### 2.2. 제안 방법

본 논문에서는 네트워크의 전체적인 처리량을 증가시키기 위해 위에서 언급한 두 가지 문제점을 해결하고자 한다. 제안 방법은 ALOHA-Q를 기반으로 하되, 수중 통신에 더 적합한 방법으로 수정하여 NAV 설정값을 학습하고 상황 내에서 최적의 값을 찾는다. 본 논문에서는 이를 Q-NAV라고 정의한다.

### 2.2.1. 적용할 수중 무선 네트워크 환경

본 논문에서는 수중 무선 네트워크에서 자주 사용되는 구조를 기반으로 한다. Fig. 2.은 수중 무선 네트워크의 구조를 간단하게 표현한 그림이다.

데이터를 수집하는 노드(Node1, 2, 3)는 송신할 데이터가 존재하는 경우 Sink Node로 데이터를 송신하고, Sink Node는 필요에 따라 데이터 가공, 지상의 기지국으로 데이터 전송 등 설정된 역할을 수행한다. Sink Node는 최대 전파 도달 범위를 가지고 있으며, 해당 범위 내의 모든 노드에 대하여 데이터를 수집한다. 노드는 서로의 위치를 알 수 없는 환경이며, 모든 수중 무선 네트워크와 동일하게 시간 동기화가 불가능한 상황을 가정한다.

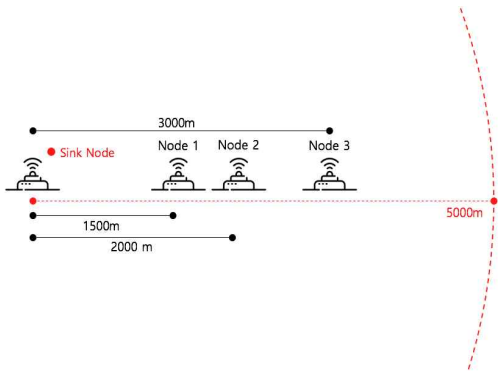

Fig. 2. Underwater wireless network example

### 2.2.2. Q-NAV

Q-NAV는 ALOHA-Q 방식을 참고하여 수중 무선 네트워크에 알맞게 수정한 방법으로, 낮은 연산 성능을 갖는 수중 노드에 적합하게 적은 연산으로 환경에 적응하는 학습을 할 수 있고 적은 에너지를 사용하는 특징을 갖는다.

Q-NAV를 적용한 수중 무선 네트워크에서의 모든 노드는 보상 테이블(Q_table)을 갖는다. 보상 테이블은 2차원 배열로 구성되며, 배열은 Sink Node와 통신 중인 노드(RTS/CTS 송신 노드)의 주소를 저장하는 공간과 RTS 또는 CTS를 수신하였을 때, NAV 선택에 대한 성공/실패 보상 값을 저장하는 공간으로 이루어진다.

Q-NAV를 적용한 수중 무선 네트워크의 각 노드는 채널 점유를 위해 전송하는 RTS/CTS 패킷 내의 NAV 값 설정에서 최대 전파 지연시간을 선택하지 않고 발송한다. 즉, 송신자는 NAV 설정 필드에 아무런 값을 넣지 않고 전송한다. NAV 값은 RTS/CTS 송신자가 아닌 RTS/CTS 수신자가 계산하는 방식을 사용한다. Q-NAV의 경우 학습을 진행하는 기본적인 방식은 다음과 같다. RTS/CTS 수신한 노드는 학습 초기의 경우, NAV값을 랜덤으로 선택하여 설정한다. NAV가 설정된 상황에서 데이터 송신이 필요하여 채널을 점유하려고 하는 경우, NAV가 끝난 직후 채널 점유를 시도한다. 채널 점유가 성공하는 경우 노드가 가지고 있는 보상 테이블에 설정한 NAV 시간에 대한 보상을 준다. 반대로 실패한 경우 패널티를 주는 방식을 사용한다. 충분한 학습이 이루어진 경우, 랜덤 방식 사용보다는 경험으로 측정된 보상 테이블을 활용하여 NAV값을 선정한다.

Fig. 3.는 Q-NAV의 작동 방식을 알고리즘화 하여 보여준다. 수중 통신 노드가 RTS/CTS를 수신 하는 경우, NAV 값을 결정하게 된다. 이때, 일정 확률로 무작위 NAV 값을 사용할지, 보상 배열의 인덱스 0에 가장 가까우면서 보상 값이 양의 수인 인덱스를 사용할지 결정한다. 이 인덱스 값은 NAV의 시간을 나타낸다. 이를 결정하는 확률 e는 통신 횟수에 따라 보정하여 사용한다. 무작위 NAV값을 사용하는 경우 값의 범위는 시스템 상의 최대 전파 지연 시간으로 계산된 NAV 값을 초과하지 않는다.

```
Q-NAV Algorithm :
if Receive_RTS/CTS():
  if Receive_Count > Threshold:
    set_e()
  Receive_Count++
  if random() < e:
    random_nav = set_random_nav_time()
    nav_end_time = set_nav_end_time(random_nav)
    waiting_nav_time()
    if communication_failure():
      Q_table[node, random_nav] = Q_table[node, random_nav] – 1
    else:
      Q_table[node, random_nav] = Q_table[node, random_nav] + 1
  else:
    LPV = selected_lowest_positive_value(Q_table[node, : ])
    nav_end_time = set_nav_end_time(LPV)
    waiting_nav_time()
    if communication_failure():
      Q_table[node, LPV] = Q_table[node, LPV] - 1
    else:
      Q_table[node, LPV] = Q_table[node, LPV] + 1
```

Fig. 3. Q-NAV Algorithm

만약 NAV 대기 시간 중 송신할 데이터가 발생하는 경우, NAV 대기 시간이 끝나는 즉시 데이터를 전송한다. 이때, 성공 여부에 따라 해당 NAV 값에 대한 보상이 이루어지며, 통신 중이던 노드 간의 거리는 알 수 없으나 두 노드 간 통신 시간은 변하지 않기 때문에 학습을 통해 해당 노드에 대한 NAV 값은 최적화된다. 즉, RTS/CTS를 송신한 모든 노드에 대한 NAV 값을 학습하고 각 노드마다 다른 NAV 값을 사용하게 된다. 또한 학습이 진행됨에 따라 e 값을 보정하여 적당한 학습이 진행된 이후에는 가장 짧으면서 양의 보상 값을 갖는 NAV 값을 선택하게 된다.

## 3. 시뮬레이션

### 3.1. Q-NAV Simulation

Fig. 4. Q-NAV Simulation parameter

본 연구에서는 Q-NAV 방식을 적용한 수중 무선 네트워크가 최적의 NAV 값을 잘 학습하는지 실험하였다. 실험은 Python 및 Python 라이브러리를 이용하였으며, 학습 시뮬레이션에 사용된 환경은 Fig. 4.와 동일한 환경을 구상하였다. 학습 방식은 Fig. 3.과 동일하고, 실제 코드에서의 구현을 위해서 필요한 부분은 수정하여 사용하였다. 실험에서는 총 4개의 노드를 사용하였고 네트워크 환경은 Fig. 2.와 동일하다.

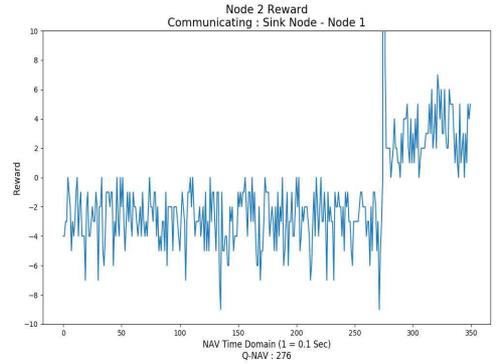

Fig. 5. Q-NAV Reward value (Node 2)

Fig. 5.는 Sink Node와 Node 1의 통신으로 발생한 RTS/CTS가 Node 2에게 전송되고, Node 2가 Q-NAV를 통해 채널 예약을 했을 때의 보상 값 그래프이다. 그래프에서 x축의 경우 NAV 값을, y축의 경우 보상 값을 의미한다. 보상 값에서 양의 수치는 통신이 성공했음을, 음의 수치는 실패했음을 의미한다.

NAV 값 0부터 275까지의 보상 값은 음수 값을 갖고 276에서 급격하게 높아졌다가 다시 낮아지는 것을 볼 수 있다. 이는 학습을 시작하고 일정 시간 이후부터는 무작위 NAV 값을 선택하지 않고 보상 배열의 인덱스가 0에 가장 가까우면서 양의 보상 값을 갖는 인덱스를 선택하기 때문이다. 이 때문에 경계선인 276을 일정 시간 이후 지속적으로 선택하게 됨으로써 매우 높은 값을 가지게 된다.

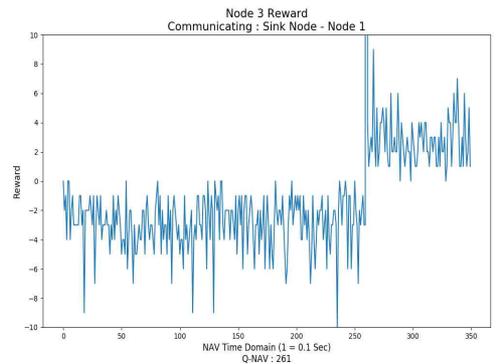

Fig. 6. Q-NAV Reward value (Node 3)

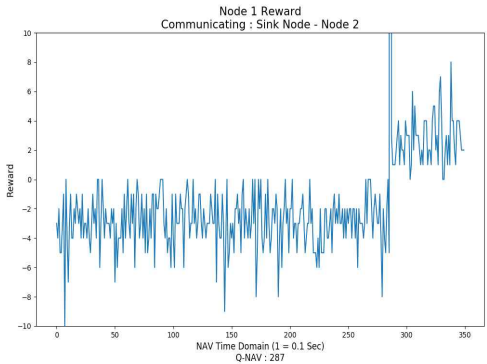

Fig. 7. Q-NAV Reward value (Node 1)

Fig. 6.와 Fig. 7.은 다른 노드 역시 모든 상황에서 학습이 잘 됨을 보여준다. Fig. 8.은 일반적인 수중 무선 네트워크(Underwater Wireless Sensor Network; UWSN)를 사용했을 때의 NAV 값, Q-NAV 시뮬레이션에서 각 Node와 Sink Node 사이에서 통신이 이루어지는 경우 해당 통신이 실제로 끝나는 시간(Real Communication End Time; RET), Q-NAV를 통해 NAV 값을 결정하여 통신을 시도하여 Sink Node에 패킷이 도착하는 시간(Learning End Time; LET)을 나타낸 표이다. LET의 경우 학습된 NAV값, 해당 노드간의 전파 지연시간, 패킷 처리 시간이 모두 계산되어 도착하는 시간을 의미한다.

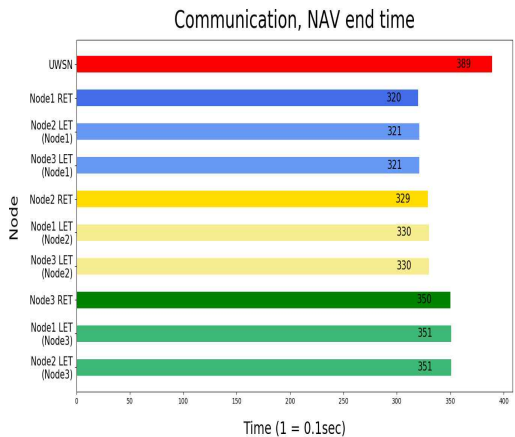

Fig. 8. Q-NAV Communication Time

일반 수중 무선 네트워크(UWSN)의 방법을 사용하는 경우 본 논문의 환경 기준으로, NAV 시간은 38.9초로 고정된다. 이는 노드 간 거리에 상관없이 매번 일정 대기 시간을 낭비하게 된다. 또한 앞서 말한 거리가 먼 노드의 통신 불공정성을 일으킨다.

Node1 RET의 경우 Node 1이 Sink Node와 통신을 하는 동안 채널을 점유하게 되는 32초를 의미한다. Node2 LET(Node1)와 Node3 LET(Node1)은 Node 1의 RTS/CTS를 수신하고 NAV 만큼 쉰 이후 데이터를 전송하였을 때, Sink Node에 패킷이 도착하는 시간을 의미한다. Node 1의 통신이 32초에 종료되는데, Q-NAV 학습을 통해 학습한 Node2와 Node3의 패킷 도착은 32초 직후인 32.1초에 도착함을 알 수 있다. 이는 패킷 전송 시 사용하는 Back-off Window를 적용하지 않은 시간으로 불필요한 대기 시간 문제를 해결했을 뿐만 아니라 노드 간 공평성 문제를 해결함을 알 수 있다. 해당 환경에서 기존 수중 무선 네트워크에서 사용하는 NAV 시간 대비, Q-NAV가 적용된 NAV 시간은 최대 6.8초 절약하는 모습을 보여주며, 이는 약 17.5%의 시간을 감소시킨 모습입니다. 노드의 위치에 따라 NAV 값이 달라지므로, 기존 대비 절약되는 성능은 항상 가변적이다.

## 4. 결론

Q-NAV는 강화 학습의 가장 큰 문제 중 하나인 많은 양의 연산을 하지 않으면서도 간단하고 확실한 학습을 보여준다. 이는 에너지 제한적인 환경과 제한적인 컴퓨팅 파워를 갖는 수중 무선 네트워크 환경에 적합한 해결법이라고 볼 수 있다. 또한 Q-NAV는 수중 무선 네트워크에서 해결해야 하는 문제인 노드 거리에 의한 통신 기회 불공정성 문제와 낭비되는 대기 시간문제를 해결함을 보여준다. 수중 무선 네트워크는 노드 간 시간 동기화가 불가능하지만 학습을 통하여 데이터 발신 타이밍을 찾는 방식을 사용하기 때문에 동기화 없이 정확히 통신이 끝나는 지점을 찾을 수 있다. Q-NAV는 기존에 수중 환경 특성상 해결하지 못하는 많은 문제를 해결하는 모습을 보여주었다. 또한 간단한 알고리즘으로 에너지 제한적인 환경과 낮은 컴퓨팅 파워를 갖는 실제 수중 네트워크에 적용이 용이할 것으로 예상된다.

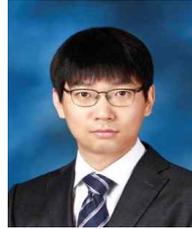

조오현(Ohyun Jo)[정회원]
- 2005년 2월 : 한국과학기술원 전기및전자공학(학사)
- 2007년 8월 : 한국과학기술원 전기및전자공학(석사)
- 2011년 2월 : 한국과학기술원 전기및전자공학(박사)
- 2011년 4월 ~ 2016년 2월 : 삼성전자 DMC 연구소
- 2016년 3월 ~ 2017년 7월 : 한국전자통신연구원
- 2017년 8월 ~ 2018년 2월 : 육군사관학교 전자공학과 조교수
- 2018년 3월 ~ 현재 : 충북대학교 소프트웨어학과 조교수

- 관심분야 : IoT 융합, 정보통신 및 네트워크, 기계학습
- E-Mail : ohyunjo@chungbuk.ac.kr

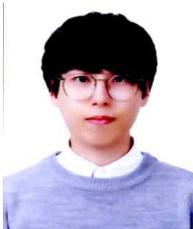

박석현(Seok-Hyeon Park)[학생회원]
- 2019년 2월 : 세명대학교 컴퓨터학과 (학사)
- 2019년 2월 ~ 현재 : 충북대학교 컴퓨터과학과 석사과정

- 관심분야 : 수중 통신, 인공지능
- E-Mail : seokhyeon@chungbuk.ac.kr